\documentclass{article}
\usepackage{graphicx}

\usepackage{amsmath}
\usepackage{amssymb}
\usepackage{float}
\usepackage{amsmath, amssymb, setspace,cite,verbatim}
\usepackage{graphicx}
\usepackage{hyperref}
\usepackage{color}
\usepackage{wrapfig}

\usepackage{multicol}
\usepackage{etoolbox}
\usepackage{relsize}
\usepackage{multirow}

\usepackage{latexsym}
\usepackage{a4wide}
\usepackage{graphicx, subfigure}
\usepackage{cite}
\usepackage{tabularx}
\usepackage{array}
\usepackage{graphics}     
\usepackage{amsmath}
\usepackage{amssymb}
\usepackage{longtable} 
\usepackage{verbatim} 
\usepackage[table]{xcolor}
\usepackage{booktabs}
\usepackage{hyperref} 
\usepackage[utf8]{inputenc}
\usepackage{soul}

\newcommand{\fms}[1]{{#1}\!\!\!\slash}

\newcommand{\nb}{\overline{n}}

\newcommand{\nbs}{\fms{\overline{n}}}

\newcommand{\braB}{\,\langle B \!\mid\,}
\newcommand{\ketB}{\,\mid\! B \rangle\,}

\begin{document}
\title{\bf \large Recent progress on the nonlocal power corrections to the inclusive penguin decays $\bar B \to X_s \gamma$ and $\bar B \to X_s \ell\ell$~\footnote{MITP-23-008. DESY-23-29. Based on talks given by T.H. at  the LF(U)V workshop, Zurich, 4.-6.7.2022 and at  the  LFC22 workshop, Trento, 29.8.-2.9.2022.} }
\date{ }

\author{ \normalsize
Michael Benzke\\ 
{\em  \normalsize Institute for Theoretical Physics, University Hamburg}\\
{\em  \normalsize Luruper Chaussee 149, D-26761 Hamburg, Germany   } \\ \\
\normalsize Tobias Hurth\\
{\em  \normalsize   PRISMA+ Cluster of Excellence and  Institute for Physics (THEP})\\
{\em  \normalsize Johannes Gutenberg University, D-55099 Mainz, Germany}
}
\maketitle

\baselineskip=10pt
\begin{abstract}
 We report  on recent progress in the field of nonlocal (so-called resolved) contributions to  the inclusive penguin decays  which presently belong to the largest uncertainties in these inclusive  decay modes. There is still a very large scale and a large charm mass dependence in the present leading order results which can in the future be  decreased  by including the $\alpha_s$ corrections. 
 \end{abstract}
\baselineskip=14pt

\newpage
 
\section{Introduction}
The inclusive decay modes $\bar B \to X_{s,d} \gamma$ and  $\bar  B \to X_{s,d}  \ell^+\ell^-$ are well known for being theoretically very  clean modes of the indirect search for new physics via flavour observables and golden modes of the Belle-II experiment  (for reviews see Refs.~\cite{Hurth:2003vb,Hurth:2010tk,Hurth:2012vp}).  The Belle-II experiment at KEK will accumulate 
two orders of magnitude larger  data samples than the $B$ factories~\cite{Belle-II:2018jsg}. This will lead to a very high experimental precision in the penguin modes which has to be  matched by the accuracy of theoretical predictions. 

Within the heavy mass expansion (HME) these  inclusive so-called penguin modes are domi- nated by the partonic contributions which can be calculated perturbatively and sub-leading contributions start at the quadratic level,                $(\Lambda/ m_b)^2$ only.  However, it is well known that this operator product expansion breaks down in these inclusive modes if one considers operators beyond the leading ones. This breakdown manifests in nonlocal power corrections, also called resolved contributions.  They are characterised by  containing  subprocesses in which the photon couples to light partons instead of connecting directly to the effective weak-interaction vertex~\cite{Lee:2006wn}. 

These resolved contributions can be systematically calculated using soft-collinear effective theory (SCET). In case of the inclusive $\bar B \to X_{s} \gamma$ decay all resolved contributions to $O(1/m_b)$ have been calculated  some time ago~\cite{Benzke:2010js,Benzke:2010tq,Lee:2006wn}. Also the analogous contributions to the inclusive $\bar B \to X_{s ,d} \ell^+ \ell^-$ decays have been analysed to $O(1/m_b)$~\cite{Hurth:2017xzf,Benzke:2017woq}. In both cases an additional uncertainty of $4-5\%$ was found which represents the largest uncertainty in the prediction of the decay rate of $\bar B \to X_{s} \gamma$~\cite{Misiak:2015xwa} and of the low-$q^2$ observables of $\bar B \to X_{s ,d} \ell^+ \ell^-$~\cite{Huber:2015sra,Huber:2019iqf}.

In both penguin decays  there are four resolved contributions at $O(1/m_b)$, namely from the interference terms ${\cal O}_{7\gamma} - {\cal O}_{8g}$,\, ${\cal O}_{8g} - {\cal O}_{8g}$, and ${\cal O}^{c}_{1} - {\cal O}_{7\gamma}$, but also from ${\cal O}^{u}_1 - {\cal O}_{7\gamma}$.~\footnote{For the definition of the operators the reader is guided to Ref.~\cite{Benzke:2017woq}} The latter is CKM suppressed,  but was shown to vanish~\cite{Benzke:2010js}. The ${\cal O}^{c}_{1} - {\cal O}_{7\gamma}$ piece is the largest contribution 
in both penguin decays.  As was already noted in Ref.~\cite{Hurth:2017xzf,Benzke:2017woq}, there are subleading contributions due to the interference of
${\cal O}_{9,10}$ and ${\cal O}^c_1$ at order $1/m_b^2$ which are numerically relevant due to the large ratio $C_{9,10} /  C_{7\gamma}$ and which have be to added in the future. 

Recently, a new theoretical input~\cite{Gunawardana:2017zix,Gunawardana:2019gep} allowed to reduce the impact of the ${\cal O}^{c}_{1} - {\cal O}_{7\gamma}$ contribution. 
However, in a more  recent   analysis of the ${\cal O}^{c}_{1} - {\cal O}_{7\gamma}$ resolved contribution  a smaller reduction  was found and additional uncertainties were identified~\cite{Benzke:2020htm}. We discuss the reasons for these discrepancies between the two analyses in  Refs.~\cite{ Gunawardana:2019gep,Benzke:2020htm} below. 

In particular, a large scale dependence and also a large charm mass dependence were identified in the lowest order result of this  resolved contribution, which calls for a systematic calculation of $\alpha_s$ corrections and renormalisation group (RG) summation~\cite{Benzke:2020htm}. For this task a factorisation formula for the subleading resolved  corrections is needed which is valid to all orders in the strong coupling constant  $\alpha_s$. Here another 
new input was given in Ref.~\cite{Hurth:2023paz} where a previous failure of factorisation in specific resolved contributions was healed by using new refactorisation techniques~\cite{Liu:2019oav,Liu:2020wbn,Beneke:2022obx}.

In this status report we focus on these two issues, namely the   factorisation theorem for resolved contributions and the estimate of the ${\cal O}^{c}_{1} - {\cal O}_{7\gamma}$ contribution.

\section{General properties of resolved contributions} 
The resolved contributions in the penguin decay $\bar B \to X_{s ,d} \ell^+ \ell^-$ were calculated in the presence of a cut in the hadronic mass $M_X$ which is needed also at the Belle-II experiment in order to suppress huge background from double semi-leptonic decays. But  it was shown~\cite{Hurth:2017xzf, Benzke:2017woq} that the resolved contributions stay nonlocal when the hadronic cut is released. Therefore they represent an irreducible uncertainty. In addition it was shown that the support properties of the shape function imply that the resolved contributions (besides the ${\cal O}_{8g} - {\cal O}_{8g}$ one) are almost cut-independent. The analogous statements for the resolved contribution in the penguin decay $\bar B \to X_{s,d} \gamma$ are also valid when the photon energy cut is moved out of the endpoint region. 

One finds a  factorisation formula for the various  contributions to the inclusive penguin 
decays~\cite{Benzke:2010js}:~\footnote{The symbol $\otimes$ denotes the convolution of the soft and jet functions.}
\begin{align}\label{fact2}
   &d\Gamma(\bar B\to X_s\,  \gamma, \ell^+ \ell^-)
   = \sum_{n=0}^\infty\,\frac{1}{m_b^n}\, \sum_i\,H_i^{(n)} J_i^{(n)}\otimes S_i^{(n)} \nonumber \\
   &\qquad + \sum_{n=1}^\infty\,\frac{1}{m_b^n}\,\bigg[ \sum_i\,H_i^{(n)} J_i^{(n)}\otimes S_i^{(n)}\otimes\bar J_i^{(n)}
    + \sum_i\,H_i^{(n)} J_i^{(n)}\otimes S_i^{(n)} \otimes\bar J_i^{(n)}\otimes\bar J_i^{(n)} \bigg] \,. 
\end{align}
The first line describes the so-called direct contributions, while the second line contains the resolved contributions. The latter appear first only at the order $1/m_b$ in the heavy-quark expansion.  
Here hard functions $H_i^{(n)}$ describe physics at the high scale $m_b$.
$J_i^{(n)}$ are the so-called jet functions which represent the physics of the hadronic final state $X_s$ at the intermediate hard-collinear scale $\sqrt{m_b \Lambda_{\rm QCD}}$. The soft functions $S_i^{(n)}$, the so-called shape functions, parametrise the hadronic physics at  the scale $\Lambda_\text{QCD}$. Within the resolved contributions we have new jet functions 
$\bar J_i^{(n)}$ due to the coupling of virtual photons with virtualities of order $\sqrt{m_b \Lambda_\text{QCD}}$ to light partons instead of the weak vertex directly.

However, the specific resolved ${O}_{8g} - {O}_{8g}$ contribution does not factorise because the convolution integral is UV divergent. The authors of Ref.~\cite{Benzke:2010js} claimed that there is an essential difference between divergent convolution integrals in power-suppressed contributions of exclusive $B$ decays and the divergent convolution integral in the present case, while the former were of IR origin, the latter divergence were of UV nature. Nevertheless, using a hard cut-off in the resolved contribution,  the sum of direct and resolved ${O}_{8g} - {O}_{8g}$ contributions  was shown to be scale and scheme independent at the lowest order. But the failure of factorisation did not allow for a consistent resummation of large logarithms.  In a recent  paper, the divergences in the resolved and in the direct contributions were identified as endpoint divergences. It was shown that also the divergence in the direct contribution can be traced back to a divergent convolution integral~\cite{Hurth:2023paz}. Recently new techniques~\cite{Liu:2019oav,Liu:2020wbn,Beneke:2022obx} were presented in specific collider applications, which allow  for an operator-level reshuffling of terms within the factorisation formula so that all endpoint divergences cancel out. This  idea of refactorisation  was now implemented in this flavour example of the resolved contributions which includes nonperturbative soft functions, the subleading shape functions,  not present in collider applications~\cite{Hurth:2023paz}. A renormalised factorisation theorem on the operator level for these resolved contributions  was derived to all orders in the strong coupling constant. This new result establishes the validity of the general factorisation theorem, given in Eq.~\ref{fact2}, - also for the  ${O}_{8g} - {O}_{8g}$ contributions. This theorem now allows for higher-order calculations of the resolved contributions and consistent summation of large logarithms~\cite{Hurth:2023paz}.

\section{Calculation of the resolved ${\cal O}^{c}_{1} - {\cal O}_{7\gamma}$ contribution}
\subsection{General stragegy}
Following the analysis in Refs.~\cite{Misiak:2015xwa}, the SM prediction for the branching ratio of ${\mathcal B}_{s \gamma}$ with a certain cut $E_0$ in the photon energy spectrum is based on the formula
\begin{equation}  \label{brB}
{\mathcal B}(\bar B \to X_s \gamma)_{E_{\gamma} > E_0}
= {\mathcal B}(\bar B \to X_c \ell \bar \nu)
\left| \frac{ V^*_{ts} V_{tb}}{V_{cb}} \right|^2 
\frac{6 \alpha_{\mathrm em}}{\pi\;C} 
\left[ P(E_0) + N(E_0) \right],
\end{equation}
where  the so-called semi-leptonic phase-space factor $C$ is determined using the Heavy Quark Effective Theory (HQET) methods~~\cite{Alberti:2014yda}.

It is important to emphasise here that the perturbative contribution $P(E_0)$ is calculated using the local operators of the electroweak effective Hamiltonian,  while the resolved contributions in the nonperturbative contribution $N(E_0)$ are calculated using SCET. Thus, scale choices and input parameters  are in principle independent of each other in both contributions.  One should keep this in mind when the uncertainty due to the resolved contributions is calculated relative to the perturbative decay rate.  As in the original analysis in  Ref.~\cite{Benzke:2010js}, the perturbative decay rate at leading order accuracy is used in the following. Moreover, the hard scale is chosen in this perturbative contribution.~\footnote{The perturbative  rate at higher orders  is often  calculated at a scale slightly smaller than the hard scale for other reasons, namely for the stabilisation of the charm mass renormalisation dependence (see i.e. Ref.~\cite{Misiak:2015xwa})}  
Because no $\alpha_s$ corrections or any RG improvements are considered in the calculation of the resolved power corrections, the scale choice is ambiguous. One first fixes the Wilson coefficients in the resolved contribution at the hard scale, but then one varies the scale of the Wilson coefficients  in the resolved contributions between the hard and the hard-collinear scale to make the scale dependence of the resolved contributions manifest.

In the following the focus will be on the most important contribution due to the interference of ${\cal O}^{c}_{1} - {\cal O}_{7\gamma}$.
Using the original notation of Ref.~\cite{Benzke:2010js} one can write this resolved contribution normalised to the perturbative leading order result as 
\begin{equation}\label{relative uncertainty}
  {\cal F}_{\rm b \to s \gamma}^{17} = \frac{C_1(\mu)\, C_{7\gamma}(\mu)}{(C_{7\gamma}(\mu_{\rm \mbox{{\tiny OPE}}}))^2}\, \frac{\Lambda_{17}(m_c^2/m_b,\mu)}{m_b}\,, 
\end{equation}
where $\mu_{\rm \mbox{{\tiny OPE}}}$ denotes the perturbative scale, $\mu$ the scale within the resolved 
contribution. At subleading power one finds~\cite{Benzke:2010js}:
\begin{equation}\label{Lambda17A}
  \Lambda_{17}\Big(\frac{m_c^2}{m_b},\mu\Big)
  = e_c\,\mbox{Re} \int_{-\infty}^\infty \frac{d\omega_1}{\omega_1} 
  \left[ 1 - F\!\left( \frac{m_c^2-i\varepsilon}{m_b\,\omega_1} \right)
  + \frac{m_b\,\omega_1}{12m_c^2} \right] h_{17}(\omega_1,\mu)\,,
\end{equation}
with the penguin function $F(x) = 4\, x\, {\rm arctan}^2(1/\sqrt{4x-1})$. The shape function $h_{17}$ is given by the following HQET matrix element:
\begin{equation}
h_{17}(\omega_1,\mu) = \int\frac{dr}{2\pi}\,e^{-i\omega_1r}\frac{\braB \bar{h} (0) \nbs i \gamma_\alpha^\perp\nb_\beta g  G^{\alpha\beta}(r\nb)h(0)\ketB}{2M_B}\,,
\end{equation}
where $n$ and $\nb$ are the light-cone vectors and $h$ and $G$ are the heavy quark and gluon field, respectively. Soft Wilson lines are suppressed in the notation. 
The variable $\omega_1$ corresponds to the soft gluon momentum. 

The general strategy to estimate the convolution integral of the perturbative jet functions and the nonperturbative shape function  $h_{17}$ is the  derivation of  general properties of the shape functions. One shows  for example PT invariance in this case which  implies that the matrix element is real. Moreover, one finds  moments of this HQET matrix element: While the zero-moment was already known  in the first analyses of the resolved contributions,  the second moment was recently derived using HQET 
techniques; moreover rough dimensional estimates of the higher  
order moments were proposed~\cite{Gunawardana:2017zix,Gunawardana:2019gep}. One can also naturally   assume 
that the support properties and the values of the soft shape function are within the hadronic range.

Besides these general properties and the estimates on the moments nothing further is known about the structure of the subleading shape functions.  Therefore, the two new analyses of the resolved  contributions~\cite{Gunawardana:2019gep,Benzke:2020htm} follow here exactly the same strategy; they 
use a complete set of basis functions, namely the Hermite functions in order to make a systematic analysis of all possible model functions fulfilling the known properties of the shape function. This systematic approach to the shape functions was  already used in several previous analyses~\cite{Ligeti:2008ac,Lee:2008xc,Bernlochner:2020jlt}. Obviously this systematic approach allows to avoid any prejudice regarding the unknown functional form of the shape functions and, thus, leads to a valid estimate of the resolved contribution. Any additional assumption calls for a clear justification.

\subsection{Numerical results}
In both new analyses~\cite{Gunawardana:2019gep,Benzke:2020htm} the maximum value of the convolution integral between jet and shape functions was found for Hermite polynomials of degree six. Higher degree polynomials do not lead to larger values. Both analyses found a  significant reduction in the values of the resolved contributions due to the 
new input of the second moment of the shape function. 

The final result for the resolved ${\cal O}^{c}_{1} - {\cal O}_{7\gamma}$ contribution in Ref.\cite{Benzke:2020htm} is:
\begin{equation}
 {\cal F}_{\rm b \to s \gamma}^{17} \in  [-0.4\%,\,4.7\%]\,,   \label{resultbsgamma}
 \end{equation}
which represents a large  reduction compared to the original estimate in Ref.\cite{Benzke:2010js}. Some comments are in order: 
\begin{itemize}
\item In the present result no $\alpha_s$ corrections are included  and no RG improvements are done. Thus, this implies a large scale dependence in our results. In the leading order result the only scale is in the
Wilson coefficients representing the hard function. Varying the LO Wilson coefficients  $C_1(\mu)\, C_{7\gamma}(\mu)$ in the resolved contribution  from the hard scale to the hard-collinear scale increases the result by more than $40\%$.
{This represents  an additional  uncertainty of the result {\it \bf not} included in Eq.\ref{resultbsgamma}}.
\item The result in Eq.~\ref{resultbsgamma} includes a large kinematical $1/m_b^2$ contribution from the ${\cal O}^c_1 - {\cal O}_{7\gamma}$ interference. One can show by inspection of all resolved $1/m_b^2$ contributions  that this  kinematical $1/m_b^2$ term is the only one  with  the same shape function of the order $1/m_b$ as in the $1/m_b^1$ term.  All other resolved $1/m_b^2$ contributions in the interference of 
${\cal O}^c_1 - {\cal O}_{7\gamma}$   include shape functions of the order  $1/m_b^2$. Those terms are not calculated  yet.  
Therefore, the signifi- cantly large $1/m_b^2$ term due to kinematical factors in the ${\cal O}^c_1 - {\cal O}_{7\gamma}$ term was included in the result~\ref{resultbsgamma} as conservative estimate of those higher order resolved contributions.
This $1/m_b^2$ term was already included in the original analysis in Ref.~\cite{Benzke:2010js}.
There it was also shown that other $1/m_b^2$ contributions  due to the interference of  ${\cal O}^c_1 -  {\cal O}_{8 g}$  and  ${\cal O}^c_1 - {\cal O}_{1}$  are  numerically negligible. 
\end{itemize}

The authors of the new analysis in Ref.~\cite{Gunawardana:2019gep} find a much larger reduction and end up with~\footnote{We translated this result to our scale fixing. The authors of Ref.~\cite{Gunawardana:2019gep} 
find ${\cal F}_{\rm b \to s \gamma}^{17} \in [-0.3\%,\,1.6\%]$ in their paper using  the hard-collinear scale in the resolved and the perturbative contribution instead of the hard scale.} 
 \begin{equation}
{\cal F}_{\rm b \to s \gamma}^{17} \in [-0.4\%,\,1.9\%]\,,
\end{equation}
There are two main reasons for this difference with the result of Ref.~\cite{Benzke:2010js},  given in Eq.~\ref{resultbsgamma}:
\begin{itemize}
\item  The charm mass dependency originates  from the anti-hard-collinear jet function $\bar J_i^{(n)}$
represented by the charm loop with a soft gluon emission. Therefore it is appropriate to use the running charm mass at the hard-collinear scale ${m}_c^{\rm MS}(\mu_{\rm hc} )$. 
The charm mass ambiguity of the charm mass was made manifest by the variation of the hard-collinear scale $\mu_{\rm hc} \sim \sqrt{m_b\,\Lambda_{\rm QCD}}$  from $1.3\, {\rm GeV}$  to   $1.7\, {\rm GeV}$ within the recent analysis in Ref.~\cite{Benzke:2010js}.  Using the present PDG value of the charm mass being ${m}_c^{\rm MS}(m_c) = (1.27 \pm 0.02)\, \text{GeV}$ and  using three-loop running with $\alpha_s(m_c) = 0.395$ and $\alpha_s(m_Z) = 0.1185$ down to the hard-collinear scale, one finds $m_c^{\rm MS}(1.5\, {\rm GeV}) = 1.19\,{\rm GeV}$ {as central value at $1.5\, {\rm GeV}$}.
The change of  the 
hard-collinear scale indicated above then leads to~\cite{Benzke:2010js}
\begin{equation}
1.14\, {\rm GeV} \leq m_c \leq 1.26\, {\rm GeV}.
\end{equation}
The parametric errors of ${m}_c^{\rm MS}(m_c)$ and $\alpha_s$ are neglected in view of the larger uncertainty due to the change of the hard-collinear scale $\mu_{\rm hc}$. \\
In the recent analysis in Ref.~\cite{Gunawardana:2019gep}, two-loop running leads to the central value  
$m_c^{\rm MS}(1.5\, {\rm GeV}) = (1.20 + 0.03)\,{\rm GeV}$. The parametric uncertainties, but no change of the hard-collinear scale is taken into account. One then finds the following variation of the charm mass
\begin{equation}
1.17\, {\rm GeV} \leq m_c \leq 1.23\, {\rm GeV}, 
\end{equation}
which was used in the analysis in Ref.~\cite{Gunawardana:2019gep}, but it is unnaturally small and, thus, represents an underestimation of the charm mass dependence.  
\item The second reason is that the authors of Ref.~\cite{Gunawardana:2019gep}  did not include any  
estimate for the $1/m_b^2$ corrections. In view of the fact that the only  resolved 
$1/m_b^2$ term with the same shape function as in the $1/m_b^1$ term is very large, this may lead to a further  underestimation of the overall uncertainty. 
\end{itemize}

Finally, it is important to note  that the local Voloshin term
is subtracted from the resolved contribution  ${\cal F}_{\rm b \to s \gamma}^{17}$. This has been done in all analyses of this specific resolved contribution to the $\bar B \to X_s \gamma$ decay rate. Therefore  this nonperturbative contribution has still to be added to the decay rate. It is given by $\Lambda_{17}^{\rm Voloshin} = (-1) (m_b \lambda_2)/(9 m_c^2)$ and corresponds to
\begin{equation} 
{\cal F}_{\rm b \to s \gamma}^{\rm Voloshin} = - \frac{C_1(\mu)\,C_{7\gamma}(\mu)\, \lambda_2} {(C_{7\gamma}(\mu_{\rm \mbox{{\tiny OPE}}}))^2\, 9\, m_c^2} = +3.3\%\,, 
\end{equation}
If one neglects the shape function effects and treats the charm quark mass as heavy one can derive the local Voloshin term from the resolved contribution ${\cal O}^{c}_{1} - {\cal O}_{7\gamma}$ 
(see section 3.2  of Ref.~\cite{Benzke:2017woq} for more details). But the local Voloshin term,  derived in 
Refs.\cite{Voloshin:1996gw,Ligeti:1997tc,Grant:1997ec,Buchalla:1997ky}, does not account for the complete resolved contribution as one can read off from the additional contribution given in Eq.~\ref{resultbsgamma}.

We close this progress report by another recent improvement of estimates of the resolved contribution.  
Based on Ref.~\cite{Misiak:2009nr}  an estimation of the resolved contribution to the ${\cal F}^{78}$ was offered in Ref.~\cite{Benzke:2010js} using  experimental data on $\Delta_{0-}$ of  the isospin asymmetry of inclusive neutral and charged $B \to X_s \gamma$ decay~\cite{Aubert:2005cua,Aubert:2007my}.   In the recent analysis~\cite{Gunawardana:2019gep}, the authors derived new bounds by taking into account  a new Belle measurement of $\Delta_{0-}$~\cite{Watanuki:2018xxg}, which leads to the experimental determination of ${\cal F}^{78}$  being the same order of magnitude as the determination using the vacuum insertion 
approximation (VIA)~\cite{Benzke:2010js} but includes the prospect to be improved by more precise data.

\section*{Acknowledgements} TH thanks the organisers of the LF(U)V workshop in Zurich and of  the  LFC22 workshop
in Trento for their generous hospitality and the excellent organisation of the workshops. TH is  supported by  the  Cluster  of  Excellence  ``Precision  Physics,  Fundamental Interactions, and Structure of Matter" (PRISMA$^+$ EXC 2118/1) funded by the German Research Foundation (DFG) within the German Excellence Strategy (Project ID 39083149).

\end{document}